\documentclass[aps,prb,twocolumn,superscriptaddress]{revtex4} 
\usepackage{graphics}
\usepackage{graphicx}
\usepackage{amsmath}
\usepackage{amssymb}
\usepackage{bm}
\usepackage{dcolumn}

\begin{document}  
\preprint{version 1.0}

   \title[]{Atomic structure and electronic properties of the cleavage InAs(110) surface}   
   \author{X. L\'opez-Lozano} 
   \affiliation{Instituto de F\'{\i}sica, Universidad Aut\'onoma de Puebla, Apartado Postal J-48, Puebla 72570,  M\'exico}  

   \author{Cecilia Noguez}
   \email[Author to whom correspondence should be addressed. Email:]{cecilia@fisica.unam.mx} 
   \affiliation{Instituto de F\'{\i}sica, Universidad Nacional Aut\'onoma de M\'exico, Apartado Postal   20-364, Distrito Federal 01000,  M\'exico}  

   \author{L. Meza-Montes} 
   \affiliation{Instituto de F\'{\i}sica, Universidad Aut\'onoma de Puebla, Apartado Postal J-48, Puebla 72570,  M\'exico}  
	  
   \date{Submitted to Physical Review B \today}  

 \begin{abstract}
The electronic properties of the cleavage InAs(110) surface are studied using a semi-empirical tight-binding method which employs an extended atomic-like basis set. We  calculate the surface electronic band structure which is studied as a function of the structural parameters of the surface. We describe and discuss the electronic character of the surface electronic states and we compare with other theoretical approaches, and with experimental observations. Finally, we discuss which atomic model better resembles  the available experimental data.
 \end{abstract}

 \pacs{73.20.At, 78.55.Cr, 78.66.Fd, 78.68.+m}
 \maketitle 

\section{Introduction}
\label{}
Low-dimensional quantum structures have shown to have unique optical and electronic properties, being important in the fabrication of new opto-electronic devices~\cite{opto}. III-V semiconductor compounds have demonstrated to be important in the development of new light emitters and detectors. In particular, Indium compounds, like InAs and InSb, are good infrared detectors due to their small bulk-band gap. Most of modern opto-electronic devices are fabricated growing monoatomic layers on a semiconductor substrate using epitaxial techniques. The epitaxial growth depends on the ambient conditions as well as in the atomic and electronic properties of the surface of the substrate. Therefore, the continuous development of semiconductor heterostructures and other structures at the nanometer scale, makes evident the need for a detailed theoretical and experimental understanding of semiconductor surfaces.

A lot of theoretical work has been done to determine the electronic properties of Gallium compounds, and their surfaces~\cite{GaAs}. On the other hand, only a few attempts have been made to characterize the InAs surfaces. The (110) surface is the natural cleavage of zincblende crystals, and it is a non-polar surface which contains equal number of cations and anions in its unit cell, showing a partly ionic bonding. In this work, we are interested in the InAs(110) surface.

Most of the theoretical~\cite{alves,Mail,beres,engels,anatoli} studies about InAs(110) do not provide a way for directly compare with experiments~\cite{and1,and2,SwanO,Swan,leed,leed1,richter,carstensen,Drube,gudat,laar}. Furthermore, the available experimental measurements are not enough to completely elucidate the atomic structure and electronic properties of InAs(110).  For example,  Andersson and collaborators~\cite{and1} found  the energies of occupied-surface states at high-symmetry points using photoemission techniques. Six years later, the same group measured again the occupied-surface states at the same high-symmetry points~\cite{and2}, founding  a systematic shift of energy of $-0.15$~eV with their previous measurements~\cite{and1}. Independently, Swantson {\it et al.}~\cite{Swan} also measured the energies of occupied-surface states at high-symmetry points and they found differences up to 0.5~eV with those reported by Andersson and collaborators~\cite{and1,and2}. In general, the interpretation of photoemission spectra has been difficult to do because of the small bulk-band gap of InAs.

Theoretically, the electronic structure and atomic positions of InAs(110) were calculated using a quantum-molecular dynamics~\cite{Mail} based in a semi-empirical tight-binding (TB) approach. Almost a decade latter, an {\em ab initio} quantum-molecular dynamics~\cite{alves} was performed. The reported atomic structure and electronic surface states differ between TB  and {\em ab initio} calculations, and also differ with the available experimental measurements. This fact is because the semi-empirical calculations were performed using an atomic reconstruction that was not fully relaxed, while {\em ab initio} calculations were performed using a DFT-LDA with a plane-wave basis set that depend on the cut-off energy to get accurate surface states, and present systematic errors to determine the precise electronic energy levels. In summary, there is no a consensus on the physical properties of the InAs(110) surface. 

In this work, we present the electronic structure of InAs(110) employing a semi-empirical TB  formalism~\cite{noguez,Vogl} and using the atomic coordinates obtained from {\em ab initio} quantum-molecular dynamics~\cite{alves}. The use of the fully relaxed slab coordinates guarantees that the calculated electronic properties include all the subtle effects of surface-induced strain and appropriate geometry. The TB approach allows us to  analyze in detail the electronic structure and compare our calculations with available experimental data. Our calculated surface electronic band structure is very similar to that calculated by Shkrebtii and collaborators~\cite{anatoli}, however, in that work they focused to study the electronic properties of the adsorption of antimony on III-V(110) surfaces, and only a rough analysis of the clean InAs(110) surface was done, and a comparison with experiments and theoretical results was not done.

\section{Theoretical method}

The III-V(110) semiconductor surfaces relax in such a way that the surface cation atom moves inwards the surface into an approximately planar configuration, with a threefold coordination with its first-neighbors anion atoms. The topmost anion atom moves outward to the surface, showing a pyramidal configuration with its three first-neighbors cation atoms~\cite{GaAs,godin}.  The geometric parameters that describe the relaxation of the surface atoms of III-V(110) semiconductor surfaces, scale linearly with the bulk lattice constant \cite{godin}. In particular, Alves {\it et al}.  found~\cite{alves} that for the InAs(110) surface the {\it pyramidal} angle at the anion, labeled by $\alpha$,  is $\sim  90^{\circ}$, the {\it in-plane} angle $\beta$ has values close to the tetrahedral bond angle $\sim 109.47^{\circ}$, and the {\it planar} angle at the cation, labeled by $\gamma$, is $ \sim 120^{\circ}$. For the ideal surface the values for $\alpha$, $\beta$ and $\gamma$ correspond to those angles of tetrahedral bonds, $109.47^{\circ}$.

The relaxed InAs(110) surface is shown in Fig.~\ref{atomos}. In Fig.~\ref{atomos}~(a) we show the top view of a surface unit cell that contains one In atom (cation), and one As atom (anion) per atomic layer. The open circles correspond to As atoms while  black circles show In atoms. The parameter $a_{0}$ is the bulk lattice constant and $d_{0}= a_{0}/2\sqrt{2}$. In Fig.~\ref{atomos}~(b) we show a side view with only the three outermost atomic layers of the slab. Here, we define the structural parameters associated to the surface relaxation whose values are given in Table~\ref{parameters}. In Fig.~\ref{atomos}~(c) we show  the corresponding Two-Dimensional Irreducible Brillouin Zone (2DIBZ).  

In our calculations, the non-polar InAs (110) surface was modeled using a slab of 50 atoms, yielding a free reconstructed surface on each face of the slab. The thickness of the slab is large enough to decouple the surface states at the top and bottom surfaces of the slab. Periodic boundary conditions were employed parallel to the surface of the slab to effectively  model an infinite two-dimensional crystal system. The atomic coordinates were taken from Ref.~3, and are given in Table~\ref{parameters}. We have performed calculations with all the structural parameters in Table~\ref{parameters}, however, those corresponding to the Density Functional Theory~\cite{alves} (DFT) with an energy cutoff of 18~Ry are the ones that best resemble the available experimental data~\cite{and1,and2,SwanO,Swan,leed,leed1,richter,carstensen,Drube,gudat,laar}. We calculate the electronic level structure of the slab using a well known parameterized TB approach with a sp$^3$s$^*$ orbital-like basis, within a first-neighbor interaction approach~\cite{Vogl}. This wave function basis provides a good description of the valence and conduction bands of cubic semiconductors. This TB approximation has been applied to calculate the electronic and optical properties of a variety of semiconductor surfaces, including other III-V compounds~\cite{noguez2}. The TB parameters are taken to be the same as those of Vogl \cite{Vogl} for the bulk but they are scaled by a factor of $(D/d)^2$, where $d$ is the bond length of any two first-neighbor atoms, and $D = \sqrt{3}a_0/4$.~\cite{harrison} These changes to the original bulk parameters provide an excellent description of the electronic structure, as compared to experimental measurements.  

The atomic structure of the surface region is intimately related to its electronic structure. Experimentally, the electronic structure can be determined by means of electron spectroscopies like photoemission (PE), inverse photoemission (IPE), and Scanning Tunneling Spectroscopy (STS). These techniques are sensitive to the surface's features and electronic properties due to reconstructions or adsorption events. We present in Section~III the surface electronic band structure and the local density of electronic states of the reconstructed or relaxed InAs(110) surface.

\section{Results}

We show the surface electronic band structure along high-symmetry points of the 2DIBZ of the reconstructed InAs(110) surface in  Fig.~\ref{ebclean}. The projected bulk electronic states are shown in tiny black dots, while the surface electronic states are shown in large black dots. The calculated Fermi energy level, ${\rm E_F}$, is at 1.1~eV above the Valence-Band-Maximum (VBM), while  the Minimum of the Conduction Band (MBC)  is at $E=0.6$~eV from the VBM at the high-symmetry point $\Gamma$. We denote the surface electronic states  using the labels $A_{i}$ and $C_{i}$ associated to the surface anions and cations, respectively, as introduced by Chelikowsky and Cohen\cite{chel}. 

Below the  VBM we found four well-defined occupied surface electronic states denoted by  $A_{5}$, $A_{3}$, $A_{2}$ and $C_{2}$.  The $A_{5}$ surface states correspond to the dangling bonds of the As atoms located at the first atomic layer. The $A_5$ states form a band from the high-symmetry point X to the point ${\rm X}'$, going through the high-symmetry point M in the 2DIBZ. This band has a minimum at X with an energy of -1.20~eV and disperse upwards towards the $\Gamma$ point. From X, the band also disperse upwards towards the M point, where the $A_{5}$ surface states have an energy of about -0.8~eV.  From M to X, this band disperses into the projected bulk band. The $A_5$ band shows a small dispersion around M given rise to a large contribution on the Local Density of States (LDOS) of the first layer at an energy of about -1~eV, as shown in Fig.~\ref{dosclean}. 

The $A_{3}$ surface electronic states are at a lower energy than $A_5$. The $A_3$ states are due to the backbonds between the anions localized in the first atomic layer and the cations in the second layer. The $A_3$ band has a minimum in the X high-symmetry point with an energy of -2.8~eV from the VBM. The band reaches its maximum at ${\rm X}'$ with an energy of -1.6~eV from  the VBM. The band shows a dispersion of 1.2~eV, however, around X and M the band is almost flat, contributing to a large density of states in the first and the second layers at energies of about -2.8~eV and -2.5~eV, respectively. This can be observed on the  LDOS  in Fig.~\ref{dosclean}, where two peaks are found at these energies in the panels showing the  LDOS in the first and the second layers. 

We found surface states with an energy of about $-6.0$~eV at the high-symmetry point X, that  form a band denoted by $C_2$. This  band shows a large dispersion of about  2.5~eV, where the minimum of the band is at M with an energy of -6.3~eV, and its maximum is around ${\rm X}'$ with an energy of -3.8~eV. These surface states are located at the cation (In) atoms and are due to the bonding between the In and As atoms at the top layer. From X to M, the band shows a small dispersion which is reflected in the LDOS where a large contribution is found at about -6.2~eV in the panel showing the projected  LDOS in the first layer in Fig.~\ref{dosclean}. From M to ${\rm X}'$, the $C_2$ band disperses upwards of about 2.5~eV given rise to a small contribution to the LDOS as shown in Fig.~\ref{dosclean}. 

At lower energies we found another occupied surface electronic band denoted  by $A_2$ which extends almost along all the high-symmetry points in the 2DIBZ. These surface states are located in the anion (As) atoms, and have a $s$ character due to  the backbonds between the atoms at the first and second layers, and some contribution is also found from the backbonds between the atoms in the second and third layers. From $\Gamma$ to M going along X, the band does not show dispersion and is at -10.2~eV. From M to $\Gamma$ going along ${\rm X}'$, the band has a dispersion of about 1~eV, showing a minimum around ${\rm X}'$. 
  
We have also found several localized states at $X$ between $\Gamma$ and $M$ with an energy from -2~eV to -1~eV. These states are inside the projected bulk band, therefore, they are resonance-like states. These resonance states, denoted by $A_4$, disperse upwards from $X$ towards both, $\Gamma$ and $M$. Most of the $A_4$ states have a $p$-character, and are localized at the anion in the first atomic layer. The $A_4$ states with lower energy, between  -1.9~eV and -2.0~eV,  show also a $s$ character and are localized at the third layer.  

Above the VBM we found two unoccupied surface states bands, namely, $C_3$ and $C_4$. The $C_3$ surface states are localized at the cations in the first and third  atomic layers. They show a strong $p$ character due to the dangling bonds at cations. At X, we found that $C_3$ has a maximum with an energy of about 2~eV, and has its minimum value between M and ${\rm X}'$ with a energy of about 1.4~eV.  Finally, at 2.7~eV from VBM we found empty surface states that form a band along all the high symmetry points in the 2DIBZ.  This band is denoted by $C_4$, and show a very small dispersion along the 2DIBZ.

\section{Discussion}

In this section we discuss our results, and compare with available experimental measurements~\cite{and1,and2,SwanO,Swan,leed,leed1,richter,carstensen,Drube,gudat,laar} and theoretical calculations~\cite{alves,Mail,beres,engels,anatoli}. The electronic properties of InAs(110) have been investigated previously using experimental techniques like photoemission (PE)~\cite{and1,and2,SwanO,Swan,richter,carstensen} and inverse photoemission (IPE)~\cite{carstensen,Drube,gudat,laar} spectroscopies. We found no agreement between experimental measurements because they present difficulties to identify the position of the  VBM or ${\rm E_F}$, and the employed samples are quite different. Furthermore, photoemission measurements are difficult to interpret since emissions from surface states are usually hidden by emissions from bulk states in InAs surfaces, due to the small bulk-band gap. 

On the other hand, theoretical calculations have been performed from {\it ab initio}~\cite{alves} and semi-empirical TB~\cite{Mail,beres,engels,anatoli} methods. We summarize our results and some of the experimental and theoretical data in Table~\ref{results}, where we show the energy values at high symmetry points in 2DIBZ of the surface electronic states denoted by $A_{5}$, $A_{3}$, $A_{4}$ and $C_{2}$. The first column shows our results, the next three columns show experimental measurements obtained by PE~\cite{and1,and2,Swan}, and the last two columns show theoretical results~\cite{alves,Mail}. The values in Table~\ref{results} are those reported in the corresponding reference, or they have been estimated from the figures in each reference, then errors of about 0.1~eV in the estimated values are expected.  

Alves {\it et al.}~\cite{alves} performed an {\it ab initio} calculation based in the Density Functional Theory (DFT) within the Local Density Approximation (LDA), where many-body effects were not taken into account. They considered slabs of only eight atomic layers (16 atoms), and the employed plane-wave basis set was expanded up to an energy cutoff of 8 and 18~Ry. They reported results for the equilibrium atomic structure and the electronic band structure. It is known that the equilibrium atomic geometries can be found with good accuracy, but underestimation and/or overestimation of  electronic states  is always present in DFT-LDA calculations due to the approximations employed, for example, usually DFT neglects many-body effects~\cite{mb1}. Furthermore, plane-wave basis expansions always present convergence problems to find localized states. Therefore, the comparison of the surface states from {\it ab initio} calculations ~\cite{alves} with semi-empirical results and PE measurements always presents deviations up to $\pm$1~eV. We also compare our results with semi-empirical TB calculations done by Mailhiot {\it et al.}~\cite{Mail}. These TB calculations employed a theoretical method similar to the one used here but with different atomic positions, that were not fully relaxed. 

Both theoretical calculations~\cite{alves,Mail} found  an empty surface state $C_3$ and was identified with dangling bond states at cations. While the DFT calculation~\cite{alves} found that $C_3$ has a minimum at X, we obtained a maximum at the same symmetry point in agreement with other semi-empirical calculations~\cite{Mail,beres,engels,anatoli}. DFT calculations reported an upwards dispersion from X of about 1.4~eV, while we found a downwards dispersion from X of about 0.6~eV. This discrepancy between DFT and semiempirical calculations is expected since DFT uses a small basis set that  can not reproduce conduction states, while semiempirical calculations with an extended basis set can do. The $C_{3}$ surface states have been measured by using inverse photoemission~\cite{carstensen,Drube,gudat,laar} but only at the X high-symmetry point. The experimental measurements assigned an energy between 1.7~eV and 1.9~eV at X, and evidence of an upwards dispersion from this point have been observed~\cite{carstensen} in agreement with our calculations. However, a more detailed experimental analysis is necessary to conclude more about empty states. 

The occupied surface states denoted by $A_5$, $A_3$, and $C_2$, were also calculated in Refs.~3 and 4. Both calculations identified the $A_5$ surface states with the dangling bond states in the anion atoms, in agreement with our results. The $A_5$ surface state found using a first-principles method~\cite{alves} is shift 0.3~eV in average, above the experimental value~\cite{and2} and our calculation. Experimentally~\cite{and2}, the $A_5$ surface state has a dispersion of about 0.15~eV from X to ${\rm X}'$ through M, while we calculate a dispersion of 0.5~eV, similar to the one obtained using DFT~\cite{alves}. The surface states denoted by $A_4$ have been also observed experimentally~\cite{and1,and2}, and calculated by Mailhiot {\it et al}.~\cite{Mail}. Like for the $A_5$ states, we can not identify $A_4$ states at the $\Gamma$ point. PE measurements showed that these states have a dispersion of about 0.55~eV from X to to ${\rm X}'$ through M, which is in agreement with our calculated value of 0.5~eV, while Mailhiot {\it et al}.~\cite{Mail} found a smaller dispersion of 0.2~eV, and this value was not calculated using DFT~\cite{alves}. Previous semiempirical results~\cite{Mail} also found the occupied states labeled by $A_4$, however, these surface states were not well identified since in their calculations the $A_4$ states show a very similar dispersion than the $A_5$ states.   In general, the $A_4$ are resonant states and they are difficult to calculate, specially if a plane-wave basis is used as the DFT calculations discussed here~\cite{alves}. Below $A_4$, it has been observed other resonant states denoted by $A_3$, using PE~\cite{and2}. The reported data for $A_3$ is quite different from our calculations and previous TB calculations~\cite{Mail}, maybe because the identification of these states is not clear experimentally~\cite{and1,and2}. On the other hand, the $C_2$ states can be experimentally identified since they are in a gap, except at the X point where they disperse into the projected bulk states.

\section{Summary}

We performed a TB calculation using a fully relaxed atomic geometry to study the electronic structure of the clean InAs(110) surface. A very detailed analysis was done, and a good agreement between our calculations and some experimental data was found. In conclusion, we found that fully relaxed atomic positions from {\it ab initio} methods in combination with our semi-empirical tight-binding calculation, better resembles photoemission measurement of cleaveage InAs(110) samples. Although the agreement between our results and other theoretical and experimental data is good, we conclude that more experimental studies are necessary to clearly elucidate the atomic relaxation and the electronic properties of InAs(110).

\acknowledgments We acknowledge the partial financial support from DGAPA-UNAM grant No.~IN104201, CONACyT grants No.~36651-E  and 36764-E.

\newpage

\begin{table} 
\begin{center} 
\vskip1cm
\begin{tabular}{|l|c|c|c|c|c|}
\hline \hline
 &$a_{0}$ (\AA)& $\Delta_{1,\perp}$ (\AA)& $\Delta_{1,x}$(\AA)&$\Delta_{2,\perp}$(\AA)& $d_{12,\perp}$(\AA) \\
\hline 
Ideal & $6.04$& $0.0$&$3/4a_{0}$ &$0.0$ &$d_{0}$  \\
DFT* & 5.844 & 0.70 & 4.656 & 0.122 & 1.463 \\
DFT**& 5.861 & 0.75 & 4.663 & 0.128 & 1.445  \\
LEED & 6.036 & 0.78 & 4.985 & 0.140 & 1.497 \\ 
\hline 
 & $d_{12,x}$(\AA)&$\omega$ (deg) & (\%)$c_{1}a_{1}$ &(\%) $c_{2}a_{1}$&(\%)$c_{1}a_{2}$\\
\hline
Ideal  & $a_{0}/2$&$0.0$ &$0.0$ &$0.0$ &$0.0$ \\
DFT*  & 3.361 & 30.7 & $-1.80$ & $-0.22$ & $-2.00$ \\
DFT**  & 3.395 & 32.0 & $-1.18$ & $-0.18$ & $-1.82$ \\
LEED & 3.597 & 36.5 & $-4.22$ & $+2.03$ & --- \\
\hline \hline
\end{tabular}
\end{center} 
\caption 
{Structural parameters as defined in Fig.~\ref{atomos}. Parameters obtained from DFT calculations using an energy  cutoff of 8~Ry (*), and 18~Ry (**), both from Ref.~3, and parameters from Low-Energy Electron Diffraction (LEED) measurements from Ref.~12.} 
\label{parameters}
\end{table}

\begin{table} 
\begin{center} 
\begin{tabular}{|l|c|c|c|c|c|c|}
\hline \hline
State & This work & PE~\cite{and1}& PE~\cite{and2}&PE~\cite{Swan}&DFT~\cite{alves}&TB$^*$~\cite{Mail}\\
\hline 
$A_{5}(\Gamma)$ &  & $-0.30$ &$-0.45$ &$-0.53$ & &$-0.3$ \\
$A_{5}(X)$ & $-1.21$ & $-1.00$ & $-1.15$ & $-0.83$ &$-0.85$ & $-0.9$\\
$A_{5}(${\rm X}'$/Y)$ & $-0.70$ & $-0.85$ & $-1.00$ & $-0.73$ & & $-0.7$\\
$A_{5}(M)$ & $-0.81$ &  & $-1.10$ & $-0.70$ &  & $-0.8$\\
\hline
$A_{3}(\Gamma)$ &  &  & $-2.60$ &  & & $-2.1$\\
$A_{3}(X)$ & $-2.81$ & $-3.1$ & $-3.25$ & $-2.72$ &$-3.21$ & $-3.1$\\
$A_{3}(${\rm X}'$/Y)$ & $-1.57$ &  & $-1.50$ &  & &$-1.4$ \\
$A_{3}(M)$ & $-2.51$ &  & $-3.40$ &  & & $-2.7$\\
\hline
$C_{2}(\Gamma)$ &  &$-3.35$ & $-3.50$ &  & & $-3.2$\\
$C_{2}(X)$ & $-6.04$ &  & & $-4.90$ & & $-5.8$ \\
$C_{2}(${\rm X}'$/Y)$ & $-3.8$ &$-3.7$  & $-3.85$ & $-3.35$ &$-3.4 \dagger$ &$-3.6$ \\
$C_{2}(M)$ & $-6.28$ & & $-6.10$ &  &$-5.46$ &$-6.1$ \\
\hline 
$A_{4}(\Gamma)$ & & &$-0.45$ & & & $-0.6$\\
$A_{4}(X)$ & $-1.53$ & $-1.6$ & $-1.75$ & & & $-1.2$\\
$A_{4}(${\rm X}'$/Y)$ & $-1.36$ & $-1.05$ & $-1.20$ & & & $-0.9$\\
$A_{4}(M)$ &$-1.03$ &  & $-1.10$ & & & $-1.0$\\
\hline \hline
\end{tabular}
\end{center} 
\caption{Experimental and theoretical values of the surface states at high-symmetry points of the 2DIBZ. The energy values are in eV, where the zero energy corresponds to the  VBM. ($\dagger$) Estimated value from Fig.~7(a) in Ref.~3. (*)~Estimated values from Fig.~9(b) in Ref.~4. An error of about 0.1~eV is expected in the estimated values.} 
\label{results}
\end{table}

\begin{figure} 
\begin{center}
\end{center}
\caption{Model of the atomic geometry of InAs(110). (a)~Top view of a surface unit cell. (b)~Side view of the first three atomic layers of the surface. (c)~Two-Dimensional Irreducible Brillouin Zone.}
\label{atomos}
\end{figure}
\begin{figure}
\begin{center}
\end{center}
\vskip-0.5cm
\caption{Electronic band structure of the reconstructed InAs(110) surface. Tiny dots represent the projected bulk states, while black dots represent surface electronic states.}
\label{ebclean}
\end{figure}
\begin{figure}
\begin{center}
\end{center}
\vskip-0.9cm
\caption{Total and the projected local density of states in the first, second and third atomic layers of reconstructed InAs(110).}
\label{dosclean}
\end{figure}

\end{document}